\documentclass[apj,twocolumn,natbib209]{emulateapj}
\usepackage{graphics}
\usepackage{natbib}
\usepackage{amsmath}

\newcommand{\am}{$^{\prime}$ }

\newcommand{\nustar}{\textit{NuSTAR}}

\begin{document}

\title{NuSTAR low energy effective area correction due to thermal blanket tear}
\author{Kristin K. Madsen$^1$,  Brian W. Grefenstette$^1$,  Sean Pike$^1$, Matteo Perri$^{2,3}$, Simonetta Puccetti$^4$, Hiromasa Miyasaka$^1$, Murray Brightman$^1$, Karl Forster$^1$, Fiona A. Harrison$^1$}
\affiliation{$^1$ Cahill Center for Astronomy and Astrophysics, California Institute of Technology, Pasadena, CA 91125, USA\\
$^2$ASI Space Science Data Center, Via del Politecnico, I-00133, Roma, Italy\\
$^3$INAF-Osservatorio Astronomico di Roma, Via Frascati 33, I-00078 Monteporzio Catone, Italy\\
$^4$ Agenzia Spaziale Italiana (ASI)-Unita' di Ricerca Scientifica, Via del Politecnico, I-00133 Roma, Italy}

\begin{abstract}
A rip in the MLI at the exit aperture of OMA, the \nustar\ optic aligned with detector focal plane module FPMA, has resulted in an increased photon flux through OMA that has manifested itself as a low energy excess. Overall, the MLI coverage has decreased by 10\%, but there is an additional time varying component, which occasionally causes the opening to increase by up to 20\%. We address the problem with a calibration update, and in this paper we describe the attributes of the problem, the implications it has on data analysis, and the solution.  
\end{abstract}


\section{Introduction}
\begin{figure}
    \centering
    \includegraphics[width=0.45\textwidth]{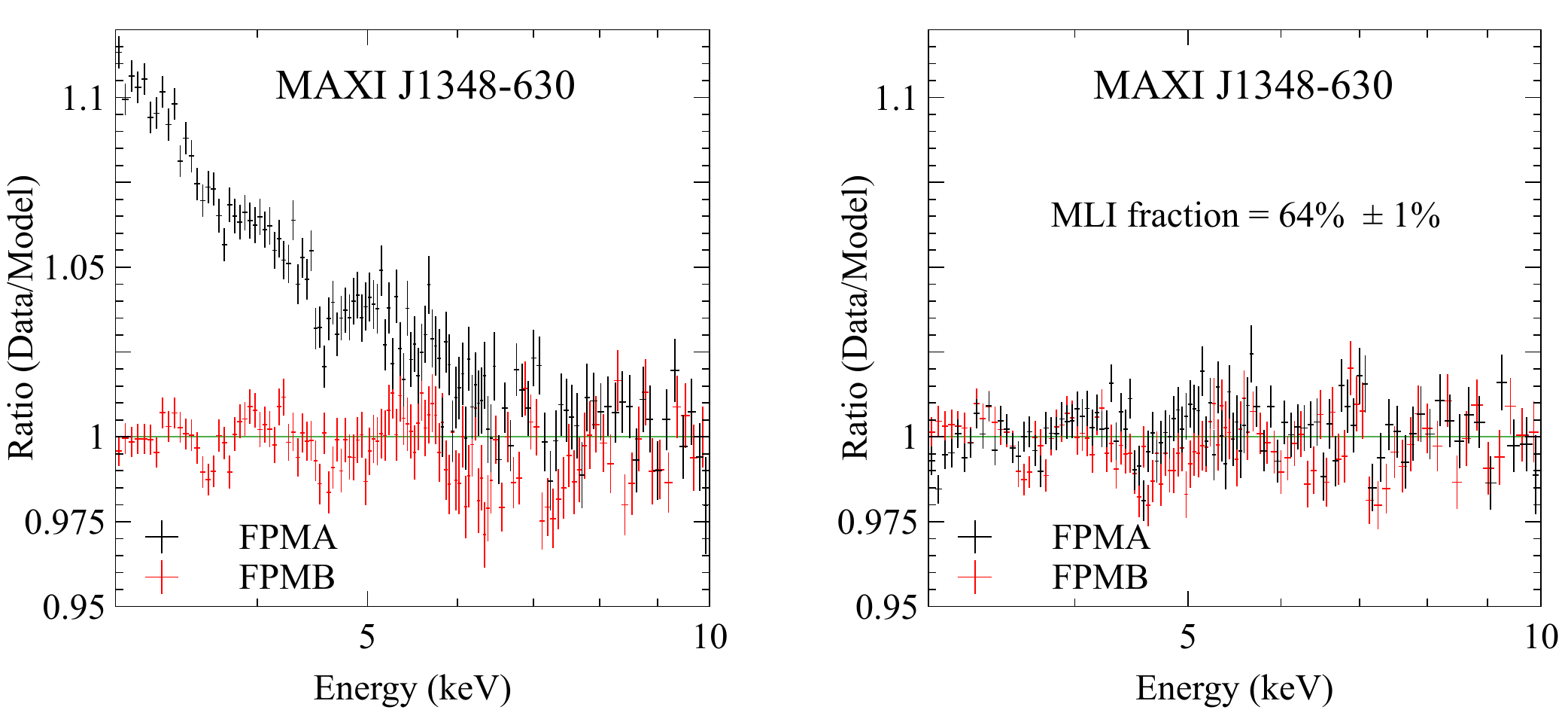}
    \caption{Left Panel: Spectra of MAXI J1348-630 (Obsid: 80402315010) fitted with the same model exhibiting a low energy excess for FPMA. Right Panel: Same spectra including a model for the MLI covering fraction where the FPMB covering fraction $\equiv$ 1.0 and the FPMA covering faction = 0.64 $\pm 0.01$.}
   \label{fig:fig1}
\end{figure}

Throughout 2019, the \nustar\ science operations center (SOC) were with increasing regularity receiving reports of differences between modules FPMA and FPMB of several percent. These differences take the form as shown in Figure \ref{fig:fig1} (left panel), where an excess of photons is seen in FPMA at low energies with respect to FPMB. Investigations into the reported cases has confirm the excess, but when compared to other sources with a similar instrument configuration, or even with the source itself at a different epoch, the excess is not always present. 

Of immediate concern, was a time dependent change of the detector and instrument response. A change in the effective area can be achieved by an unknown tilt of the optic, which would move the optical axis to a different location than presumed by the original calibration. We investigated this possibility and rejected it on the grounds that only the low energy spectrum shows deviations, while the high energy spectrum between FPMA and FPMB hasn't changed, as would have been expected due to the mirror response being more sensitive with off-axis angle at increasing energies. We also considered the possibility that one of the detector responses had changed. Fortuitously, however, we observed the excess during a Crab calibration observation in which we also obtained a stray light observation (see \cite{Madsen2017a} and \cite{Madsen2017b} for details on the stray light), and were able to confirm that in the stray light observation, which excludes the optics, there were no differences in the FPMA and FPMB spectra.

This narrowed down the source of the excess to be associated with the static absorption elements encountered by the photon on its way from the optic to the detector. The apparent variations as a function of time could be a geometric effect, as in the way the source illuminates the detector array and entrance apertures. jmhbgWe investigated the known variations in the detector absorption dead layer (see details on dead layer in \citet{Madsen2015}), and while significant, it could be ruled out since sources illuminating the same patch on the detector were observed to have a different excess.

This left two possibilities: 1) the Multi Layer Insulation (MLI), which encloses the optics, or 2) the Be window located at the entrance of the focal module housing. A hole in the Be window would have caused noise in the detectors, and since this has not been seen, the MLI remains the only plausible source.

\section{Behavior of the MLI rip}
The conjecture that the MLI is responsible for the increased flux in FPMA is supported by the fact that Optics Module A (OMA matched to FPMA) has for years been suffering from a rip in the MLI.
In August 2017, we began observing increased temperature swings in OMA, which indicated that part of the backside of the optic (facing the detector) was being exposed to space. We interpreted this as a rip in the MLI cover. In January 2019, we began to see the effect of the rip also in another temperature sensor on the backside of the optics which suggested the rip had enlarged. Combining this evidence with the low energy excess, we concluded that the most likely explanation for the excess is that the rip has become large enough to allow more photons through to the focal plane.

To test this hypothesis, we constructed an XSPEC mtable to fit the fraction, $F$, of the MLI covering the optic. At the affected energies, it is fair to assume (confirmed by raytracing) that the number of photons exiting the optics is proportional to the area and not strongly affected by vignetting. A $F = 1$ means the MLI fully covers the opening, and $F = 0.9$ that only 90\% of the opening is covered by the MLI. Figure \ref{fig:fig1} (right panel) shows MAXI~J1348-630 (ObsID 80402315010) without and with the corrected MLI, yielding a 65\% covered fraction of the aperture on the backside of the optic.

The next step was to investigate why only some observations showed the excess. To research the dependency on time, we constructed a simple method to quantify the excess. We multiplied the effective area ($A_{eff}$) with the detector efficiency ($Det_{eff}$) and divided the source counts with this simplified response to remove the instrument effect, then divided the two modules with one another and scaled by the exposure time, which may differ between modules, to remove the effect of the source spectrum:
\begin{equation*}
\begin{aligned}
    &FPMA/FPMB = \\
    &\frac{counts_A}{counts_B} \times \frac{A_{Eff}(FPMB)\circ Det_{eff}(FPMB)}{A_{Eff}(FPMA) \circ Det_{eff}(FPMA)} \times \frac{exp_B}{exp_A}.
\end{aligned}
\end{equation*}

At energies below 10~keV the detector response is largely diagonal, and the ratio mostly removes the impact of source spectrum, which for the purpose of measuring the MLI covering fraction is accurate enough. We can therefore fit the ratio directly with the MLI fraction without having to deal with the source spectrum, and performed this operation on the entire \nustar\ data library for all sources above 3 count second$^{-1}$ and within 2.5 arcminutes of the optical axis, with the additional restriction that all counts in the extraction region had to be located on Detector 0. We then correlated this data with the average temperature of OMA for each observation, and Figure \ref{fig:fig4} shows as a function of time and temperature the evolution of the rip. The plot illustrates several points:
\begin{enumerate}
\item the accuracy of the MLI fraction is on the order of $\pm 5\%$, due to other cross-calibration effects between FPMA and FPMB, such as the optical axis location, 
\item the FPMA/FPMB difference has since launch been at about 0.96, indicating that we did not have the pre-launch thickness of the FPMA (or FPMB) correct,
\item although the temperature sensors only began to detect the rip in mid 2017, it was already likely present in early 2017, 
\item the propagation of the rip appears to have stopped in 2018, 
\item however, since 2019 we occasionally see very large differences in FPMA/FPMB correlated with a very cold temperature of the optic.
\end{enumerate}

The data show that under certain circumstances the rip is apparently able to increase in size, and that when this happens the sensor measures a low temperature, presumably because more of the optic is exposed to space. We verified this with a data set of MAXI~J1348-630 taken during a period when the temperature changed in the middle of the observation, and shown in Figure \ref{fig:fig3} is the temperature in the lower panel and associated FPMA/FPMB ratios for the two epochs in the upper. This confirms that the fraction of MLI in the photon path is correlated with temperature, and indicates that the rip in the MLI is able to enlarge, resulting in a decreased temperature and increased photon flux. 

Mechanically, it is difficult to explain how the rip in the MLI can expand and contract, however, the change in the temperature is confirmed through multiple OMA temperature sensors, ruling out the possibility that the low temperature readings are due to a single faulty thermistor.

\begin{figure}
    \centering
    \includegraphics[width=0.45\textwidth]{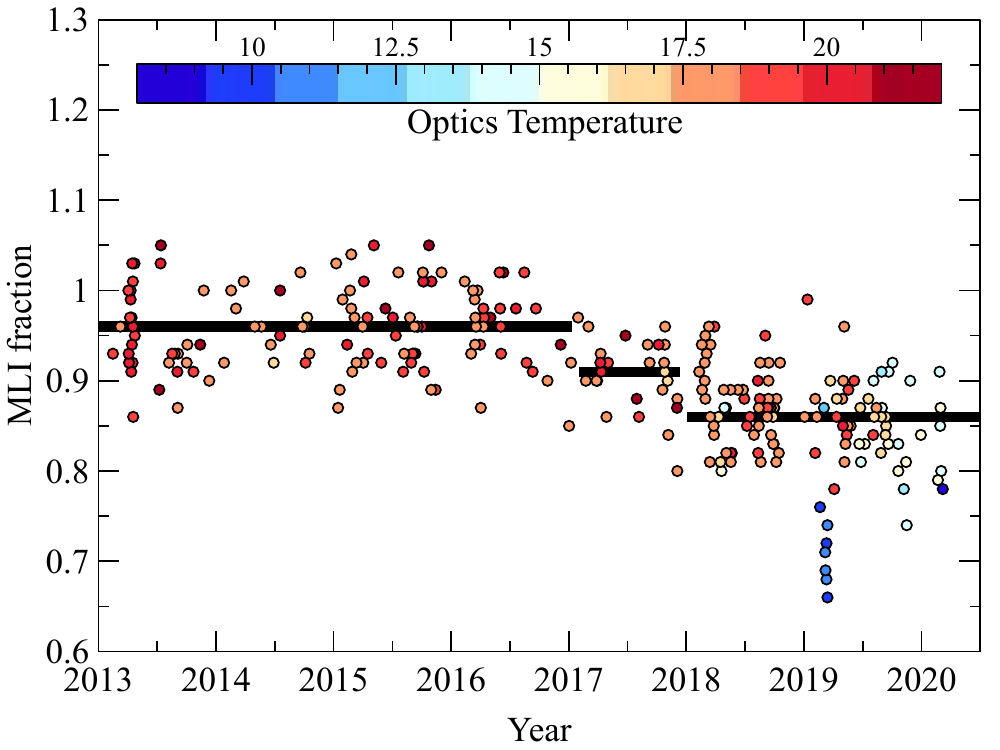}
    \caption{The covering fraction of MLI as a function of time color coded with the temperature of OMA sensor OPT0\_5TEMP. The observations are used are from the entire \nustar\ flight library fulfilling the following requirements: average count rate greater then 3 counts s$^-1$, average off-axis angle $<$ 2.5\am, and extraction region entirely on Det 0. Solid lines show mark the average level of the MLI fraction for the given ranges.}
   \label{fig:fig2}
\end{figure}

\begin{figure}
    \centering
    \includegraphics[width=0.45\textwidth]{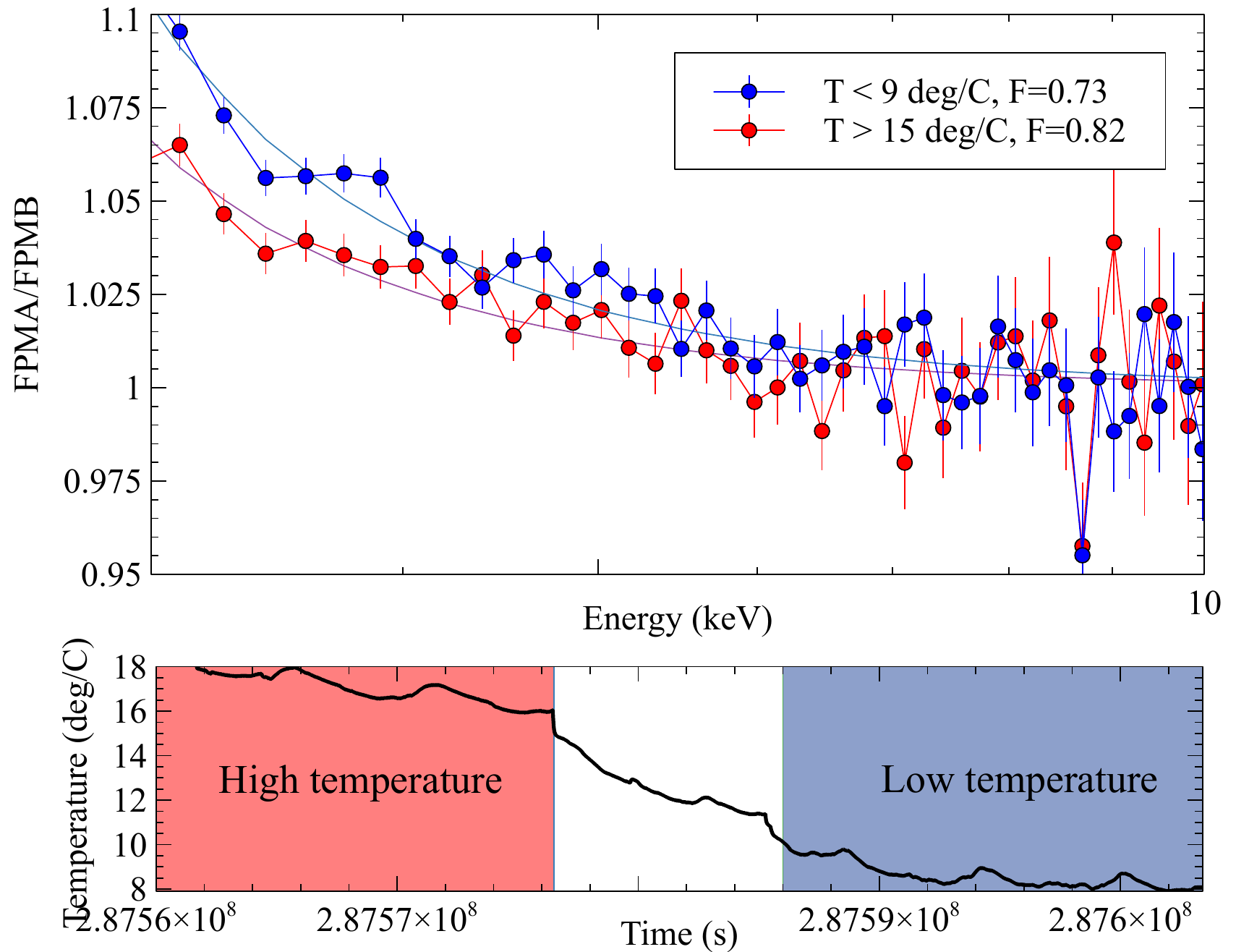}
    \caption{Two epochs of the MAXI J1348-630 (ObsID: 80402315008) FPMA/FPMB ratios. During this observation the temperature changed drastically, indicating the MLI `opened up'. The \textit{high} temperature spectrum spectrum (red) had temperature above 15 deg/C and the \textit{low} temperature spectrum (blue) had a temperature below 9 deg/C. The difference in the MLI covering fraction was 0.82 for high and 0.73 for low.}
   \label{fig:fig3}
\end{figure}

\section{The Solution}
We can correct the issue by decreasing the fraction of MLI covering the backside of the optics in the instrument response files. We can track the decrease of the MLI as a function of time by assuming the rip occurred in 2017 and that it stabilized in 2018. Taking the mean value of the data set described above for observations before 2017, we find an average MLI coverage of 0.96. Because this happened before the rip, we interpret this to mean that we overestimated the pre-launch thickness of the MLI by 4\% on OMA, not to be confused with the covering fraction after the rip. For the observations after 2018 we find an average MLI fraction of 0.86, showing that the tear has reduced the MLI coverage by $\sim $10\%. To correct for the difference, we have adjusted the on-axis ARF of FPMA by decreasing the \textit{thickness} of the MLI to F=0.96 before 2017, and changing the \textit{covering fraction} by F=0.91 in 2017 and F=0.86 after 2018. See Figure \ref{fig:fig2} and Table \ref{table:mli} for the valid ranges.

\begin{table}
\centering
\caption{Time dependent responses}
\label{table:mli}
\begin{tabular}{|l|l|}
\hline
Date range & MLI fraction \\
\hline
date $<$ 2017 & 0.96 \\
2017 $\le$ date $<$ 2018  & 0.91 \\
2018 $\le$ date $<$ 2019 & 0.86 \\
\hline
\end{tabular}
\end{table}

\begin{figure}
    \centering
    \includegraphics[width=0.45\textwidth]{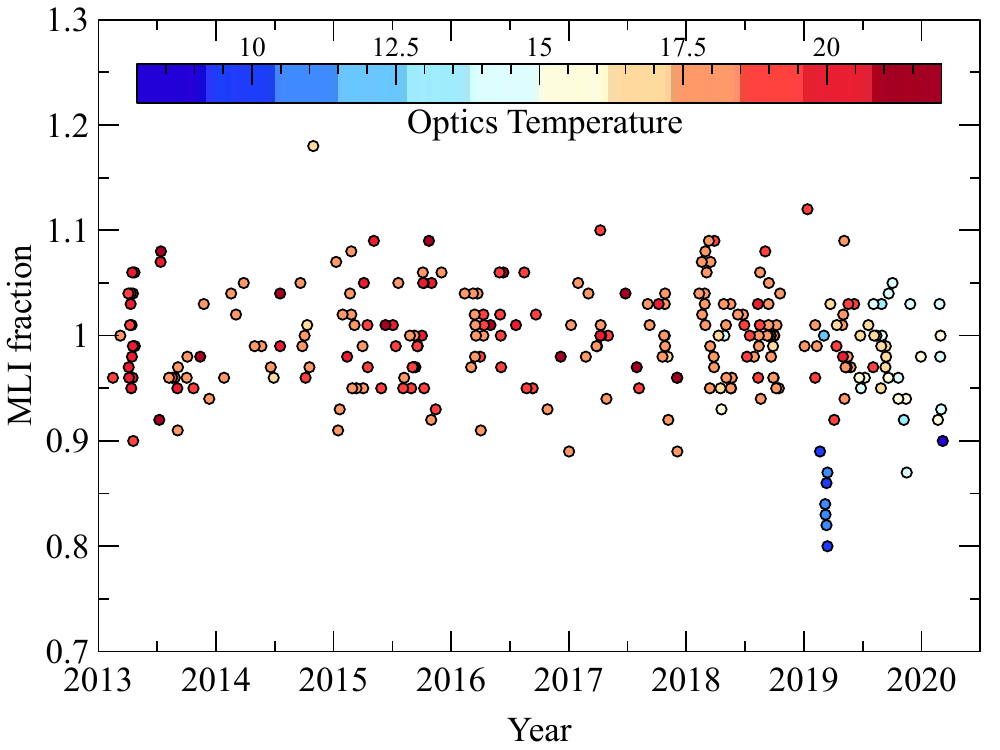}
    \caption{The covering fraction of MLI as a function of time color coded with the temperature of OMA sensor OPT0\_5TEMP. This has been adjusted with the average MLI covering fraction from Table \ref{table:mli}.}
   \label{fig:fig4}
\end{figure}

We applied these new ARFs to the \nustar\ data and recalculated the MLI fractions as shown in Figure \ref{fig:fig4}. 

\begin{figure}
    \centering
    \includegraphics[width=0.45\textwidth]{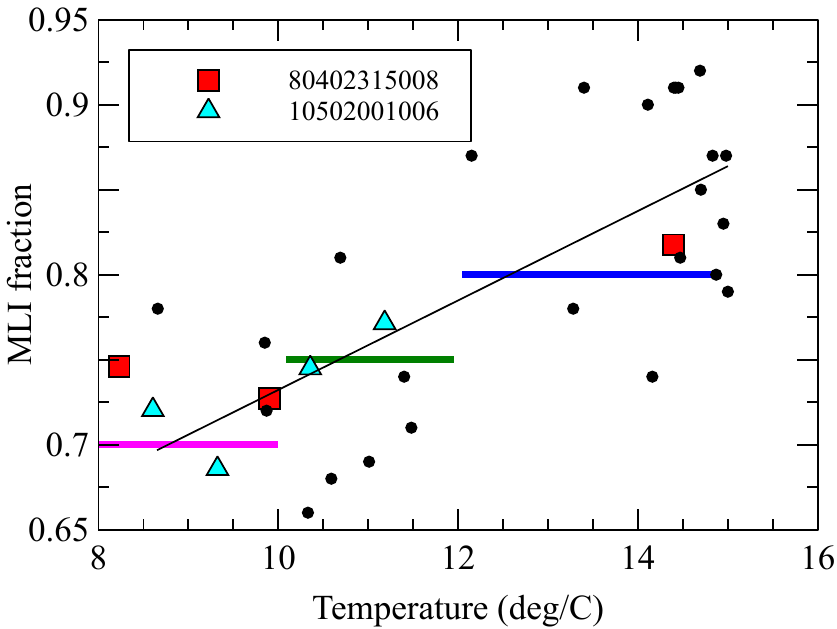}
    \caption{MLI covering fraction as a function of temperature. Black dots are observations with mean OPT0\_5TEMP $\le$ 15 deg/C and solid line is the linear fit to the points. Red squares: spectra of ObsID 80402315008 filtered into temperature bins and fitted for the MLI covering fraction (see Figure \ref{fig:fig3}). Cyan triangles: spectra of ObsID 10502001006 filtered into temperature bins and fitted for the MLI covering fraction. Colored horizontal lines show the MLI covering fraction assigned to the temperature range.}
   \label{fig:fig5}
\end{figure}

The additional time varying component, caused by the expansion of the rip, can be correlated with temperature as shown in Figure \ref{fig:fig5}. The plot shows in black circles the average temperature of all observations (above 3 counts s$^{-1}$, 2.5\am on-axis, and on Detector 0) together with a line fitted through the points. Additionally, we show in triangles MAXI~J1348-630 (obsID 80402315008) divided into temperature bins and fitted with the MLI fraction, and in squares the Crab (obsID 10502001006) also divided into temperature bins and fitted with the MLI fraction. The data shows there is a spread when using the average temperature during an observations, but that when the measurement of the MLI fraction is used with more precise temperature data then the points fall close to the line fitted through the average temperature data. Based on this we picked three temperature bands and assigned the constant MLI fractions as shown in Table \ref{table:temperature} to each band. 

\begin{table}
\centering
\caption{Temperature dependent responses \\ (for data taken after 2019-01-01)}
\label{table:temperature}
\begin{tabular}{|l|l|}
\hline
Temperature range & MLI fraction \\
\hline
$ T > 15$ & 0.86 \\
$12 < T \le 15$ & 0.8 \\
$10 < T \le 12$ & 0.75 \\
$T \le 10$ & 0.7 \\
\hline
\end{tabular}
\end{table}

These corrections only apply if the temperature is below 15 deg/C. Otherwise the global corrections discussed above apply.

\begin{figure}
    \centering
    \includegraphics[width=0.45\textwidth]{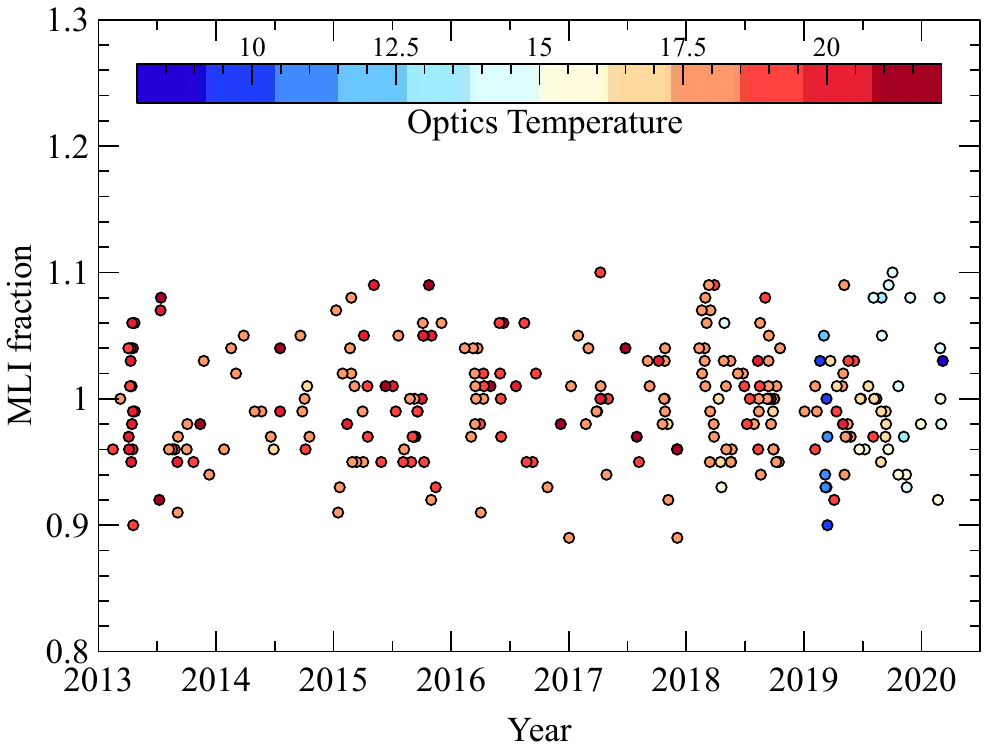}
    \caption{The covering fraction of MLI as a function of time color coded with the temperature of OMA sensor OPT0\_5TEMP. This has been adjusted with the average MLI covering fraction from Table \ref{table:mli} and \ref{table:temperature}.}
   \label{fig:fig6}
\end{figure}

When applying the corrections from Table \ref{table:mli} and \ref{table:temperature} to the \nustar\ library we see in Figure \ref{fig:fig6} that the cold outliers are now brought into family.  

The corrections are not perfect and there may be individual cases for which one may still measure an FPMA/FPMB difference. For those cases we supply the XSPEC \texttt{mtable}, however, care must be taken when using it. The exponential shape of the MLI correction is unfortunately degenerate with many astrophysical models, and we strongly advise using the following steps to minimize cross-talk with other continuum models:
\begin{enumerate}
    \item First ignore FPMA between 3 -- 7~keV, when doing your original fits. Make sure the model  includes a constant between FPMA and FPMB. Find the value of the constant and freeze it.
    \item Notice FPMA 3 -- 7~keV and add the \texttt{mtable} to the model. Set the fraction value of the MLI to 1.0 for FPMB and freeze it. Allow only the MLI fraction of FPMA to fit. 
    \item Please note that the range of MLI fraction should now lie between 0.9 and 1.0, and if the range is outside that, something else may be wrong. 
\end{enumerate}

If these steps are not followed, and the mtable parameters and the constants between A and B are allowed to fit freely, incorrect estimates of the continuum and line shapes may occur.

\section{\texttt{mtable} and {\it numkarf}/CALDB releases}
The XSPEC \texttt{mtable} is available at the \nustar\ SOC webpages \footnote{www.srl.caltech.edu/NuSTAR\_Public/NuSTAROperationSite/mli.php}. It should only be used with consideration to the caveats discussed above. It should not be used to force a model into agreement, but only if significant differences are observed between FPMA and FPMB.

An updated version of the {\it numkarf} module is included in  NuSTARDAS v2.0.0 included in HEAsoft release 6.28. The  module was modified to take into account the time and temperature dependency of the effective area of the module FPMA. This step has required the release of an associated new CALDB version (20200811). For data taken before 2019-01-01, {\it numkarf} applies a time dependent correction to  the effective area using specific on-axis ARF files as detailed in Table \ref{table:mli}. For data taken after 2019-01-01, {\it numkarf} applies a temperature dependent correction of the effective area (see Table \ref{table:temperature}). 

{\it numkarf} first reads the temperature of the OMA from the 'OPT0$\_$5TEMP' column of the input OBEB Housekeeping FITS File, calculates its average value during the observation and then retrieves the appropriate ARF file from CALDB. The new 20200811 CALDB release contains a set of new on-axis ARF files which depend on the temperature of the OMA. 

\section{Conclusion}
We have presented in this paper an important change and analysis update to \nustar\ that must be considered for all observations. The issue concerns an excess in low the energy photons in FPMA with respect to FPMB, and stems from a rip in the MLI cover of FPMA. We have released a new CALDB (version 20200811) and a new version of the software (NuSTARDAS v.2.0.0) to adjust the FPMA response to account for the reduced amount of MLI absorption. These updates include the time and temperature dependent corrections for the ARF files and will correct most data. However, where may be a subset of observations, particularly bright ones, which may not entirely be corrected by this, and we have in addition provided an \texttt{mtable} to be used for those observations. Care must be taken in applying this \texttt{mtable} as it may potentially skew the results if not used properly.

\acknowledgments
This work was supported under NASA Contract No.
NNG08FD60C, and made use of data from the NuSTAR mission,
a project led by the California Institute of Technology,
managed by the Jet Propulsion Laboratory, and funded by the
National Aeronautics and Space Administration. We thank
the NuSTAR Operations, Software and Calibration teams for
support with the execution and analysis of these observations.
This research has made use of the NuSTAR Data Analysis
Software (NuSTARDAS) jointly developed by the ASI Space Science
Data Center (SSDC, Italy) and the California Institute
of Technology (USA).

{\it Facility:} \facility{NuSTAR}

\bibliography{bib}
\bibliographystyle{jwapjbib}

\end{document}